\documentclass[12pt]{iopart}
\usepackage{graphicx,epsfig}
\usepackage{iopams,amssymb,bm}  
\begin{document}
\title{Ripples in a graphene membrane coupled to Glauber spins}
\markboth{L L Bonilla and A Carpio}{Ripples in a graphene membrane coupled to Glauber spins}
\author{L L Bonilla$^{1,3}$ and A Carpio$^{2,3}$ }
\address{$^1$G. Mill\'an Institute, Fluid Dynamics, Nanoscience and Industrial
Mathematics, Universidad Carlos III de Madrid, Avda.\ Universidad 30; E-28911 Legan\'es, Spain}
\address{$^2$Departmento de Matem\'atica Aplicada, Universidad Complutense de Madrid; E-28040 Madrid, Spain}
\address{$^3$School of Engineering and Applied Sciences, Harvard University, 29 Oxford Street, Cambridge, Massachusetts 02138, USA}
\eads{\mailto{bonilla@ing.uc3m.es},\mailto{carpio@mat.ucm.es}}
\date{\today}
\begin{abstract}
 We propose a theory of ripples in suspended graphene sheets based on two-dimensional elasticity equations that are made discrete on the honeycomb lattice and then periodized. At each point carbon atoms are coupled to Ising spins whose values indicate the atoms local trend to move vertically off-plane. The Ising spins are in contact with a thermal bath and evolve according to Glauber dynamics. In the limit of slow spin flip compared to membrane vibrations, ripples with no preferred orientation appear as long-lived metastable states for any temperature. Numerical solutions confirm this picture. 
\end{abstract}
%61.48.Gh: structure of graphene, 68.65.Pq: graphene films, 64.70.-p: specific phase transitions, 64.60.De: Statistical mechanics of model systems (Ising model, Potts model, field-theory models, Monte Carlo techniques, etc.); 05.40.-a: Fluctuation phenomena, random processes, noise, and Brownian motion; 05.45.-a: Nonlinear dynamics and chaos
%\pacs{05.40.-a; 64.60.De; 05.45.-a} %
%\noindent{\it Keywords\/}: stochastic particle dynamics (theory), stochastic processes (theory), classical phase transitions (theory)

\maketitle %\vspace{-1.2in}

\section{Introduction}
\label{sec:1}
The experimental discovery of graphene (single monolayers of graphite) \cite{nov04,nov05} and its extraordinary properties have fostered an enormous literature reviewed by a number of authors \cite{gei07,gei09,cas09,voz10}. The first observations of suspended graphene sheets showed evidence of ripples: nanometric ondulations of the sheet with characteristic amplitudes and wave lengths that do not have a preferred direction \cite{mey07,ban09}. Ripples are expected to have a significant impact on electronic transport in graphene \cite{gei08}. There have been some attempts to explain rippling as an instability or an equilibrium phase transition. For example, at zero temperature, electron-phonon coupling may drive the graphene sheet into a quantum critical point characterized by the vanishing of the bending rigidity of the membrane \cite{sgg11}. Ripples arise then due to spontaneous symmetry breaking in a buckling transition \cite{sgg11}. The connection between rippling and electronic properties was also investigated earlier \cite{kim08,gaz09}. Monte Carlo calculations also showed ripple formation at different temperatures and evidence that ripples may be connected to variable length $\sigma$ bonds of carbon atoms \cite{fas07}. The possibilities that the ripples are due to thermal fluctuations \cite{fas07,abe07} or to adsorbed OH molecules on random sites \cite{tho09} have also been explored. There is active research about the effects of ripples and strain on electronic properties, including possible strain engineering \cite{gui09}; see the review in \cite{voz10}. 

Recent experiments have shown both micron long ripples on graphene \cite{bao09} and very short \AA ngstrom-sized ripples in bilayer graphene \cite{mao11}. On large graphene sheets suspended on substrate trenches Bao et al have shown that long ripples along the direction of applied stress can be thermally induced and controlled by thermal cycling in a clean atmosphere \cite{bao09}. In contrast to \cite{mey07}, rectangular graphene sheets on substrate trenches are clamped only on two of their sides and free on the other two\cite{bao09}. The resulting ripples are much longer (wavelength about 5 microns) than those observed on samples pinned on all their sides \cite{mey07,ban09}. Bao et al interpret their experimental results by means of a theory of wrinkling in thin elastic sheets based on simplified stationary F\"oppl-von Karman equations \cite{cer03}. A different question is which ripple solutions are selected by the dynamics in the sheet. Our numerical simulations of the time-dependent von Karman plate equations (discretized on the honeycomb lattice) show that a suspended sheet that is initially in a rippled state remains rippled whereas a flat sheet remains flat. In the absence of longitudinal stretching, the initial condition selects whether the graphene sheet remains flat or it acquires curvature or ripples. 
 
In bilayer graphene, \AA ngstrom-sized ripples seem to be caused by the preparation of the sheet. Some solvent gets trapped between the two layers of the graphene bilayers and the stresses arising during drying provoke these extremely short ripples \cite{mao11}. This view is supported by the fact that these ripples do not appear in similarly prepared monolayer graphene \cite{mao11}. \AA ngstrom-sized ripples can be described by a two-dimensional (2D) Ising model with antiferromagnetic couplings involving nearest and next-nearest neighbors \cite{oha12}. The Ising spins of \cite{oha12} are not coupled to the membrane displacement vector at each site.

In this paper, we explain the small-scale ripples observed in suspended monolayer graphene sheets \cite{mey07,ban09} differently. Firstly, observation of ripples with a high-resolution transmission electron microscope (HRTEM) implies that the graphene sample is bombarded by a low-intensity electron beam. Even though the electron beam does not have sufficient energy to knock off carbon atoms, it may create defects by inducing bond rotation and certainly excite the atoms. Thus an observed graphene sample is continuously excited and cannot be considered to be in equilibrium. We suggest that the electron beam may push the carbon atoms vertically away from the planar configuration of the graphene sheet in a random fashion. As a possible model we consider a membrane with atoms coupled to Ising spins whose sign indicates the atoms trend to move upward or downward from the planar configuration. These Ising pseudo spins do not correspond to true spins but they exert forces at each atom site on the graphene sheet and interact with each other through their coupling to the membrane. Related Ising spin-membrane models with antiferromagnetic coupling have been proposed to explain wrinkling in membranes associated with partially polymerized vesicles \cite{CN93}. Elasticity coupled differently to a soft spin has been used to study two-dimensional melting mediated by dislocations \cite{CN96}. In a simpler context, we have found out that one-dimensional mechanical systems coupled to Ising spins that flip randomly according to Glauber dynamics \cite{Gl63} have a buckling equilibrium phase transition at a critical temperature that increases with sample size. More importantly, these systems exhibit ripples at any temperature in the limit as the spin relaxation time is much longer than the vibration periods of the mechanical system \cite{bcpr11}. The ripples are metastable quasi-equilibrium states about which the mechanical system experiences small and rapid oscillations. 

In the case of suspended graphene sheets, we can describe the position of carbon atoms by discrete elasticity equations previously used to model defect dynamics in graphene \cite{CBJV08,car08,bon11,BC11}. Neighboring carbon atoms in a graphene sheet attach to each other using three of their bonds. The fourth bond is not saturated and, similarly to the effect of the electron beam, may try to pull the atom upward or downward from the flat sheet configuration. As we indicated before, this trend is modeled by coupling each atom to an Ising spin. These spins are in contact with a thermal bath and flip randomly according to Glauber dynamics. After a transient stage, a random initial spin configuration gives rise to a number of spin-up and spin-down domains that, in turn, produce domains of atoms displaced upward or downward from the planar configuration. These randomly oriented domains form the ripples. Once they appear, the ripples change slowly through spin flips of the atoms in their boundaries followed by changes in their vertical displacement.  

The rest of the paper is as follows. The model we use is described in Section \ref{sec:2}. In Section \ref{sec:3}, we show that the model has a buckling equilibrium phase transition at a critical temperature $\theta_c$ which turns out to be many times higher than room temperature for the samples we consider. For $\theta>\theta_c$, the spins are unpolarized and the planar graphene sheet is thermodynamically stable. Below $\theta_c$, the spins acquire a finite polarization and the graphene sheet buckles. This phase transition corresponds to a bifurcation of the macroscopic equations of motion for mean values of the elastic displacements and the spin polarization. Since TEM observation results in a low-intensity but continuous bombardment of the graphene sheet by electrons, true thermodynamic equilibrium is never reached. In Section \ref{sec:4}, we discretize the equations of motion on the honeycomb lattice, write them in primitive coordinates and periodize them, so that integer multiples of displacements on primitive directions leave the lattice unchanged. We then solve numerically the resulting stochastic equations of motion in the limit of slow spin flipping and fast membrane vibrations. We show that long-lived ripples with no preferred orientation emerge after a short transient even from random initial conditions. These ripples correspond to those observed in experiments \cite{mey07,ban09}. We also study how the presence of defects in graphene affects rippling. Lastly, Section \ref{sec:5} contains our conclusions.

\section{Model} 
\label{sec:2}
In the continuum limit, elastic deformations of graphene sheets have the free energy of a membrane 
\begin{eqnarray}
&& F_g= \frac{1}{2}\int [\kappa(\nabla^2w)^2 + (\lambda u_{ii}^2 + 2\mu u_{ik}^2)]\, dx\, dy,  \label{eq1}\\
&& u_{ik}= \frac{1}{2}(\partial_{x_k}u_{i}+\partial_{x_i} u_k+\partial_{x_i}w\partial_{x_k}w),\, i,k=1,2,  \label{eq2}
\end{eqnarray}
where $(u_1,u_2)=(u(x,y),v(x,y))$, $w(x,y)$, $\kappa$, $\lambda$ and $\mu$ are the in-plane displacement vector, the membrane vertical deflection, the bending stiffness (measured in units of energy) and the 2D Lam\'e coefficients of graphene, respectively \cite{nel02}. $\nabla=(\partial_x,\partial_y)$ is the 2D gradient and $\nabla^2$ the 2D laplacian. In (\ref{eq2}) we have ignored the small in-plane nonlinear terms $\partial_{x_i}u\partial_{x_k}u+\partial_{x_i}v\partial_{x_k}v$. 

Carbon atoms in the graphene sheet have $\sigma$ bond orbitals constructed from $sp^2$ hybrid states oriented in the direction of the bond that accommodate three electrons per atom. The remaining electrons go to $p$ states oriented perpendicularly to the sheet. These orbitals bind covalently with neighboring atoms and form a narrow $\pi$ band that is half-filled. The presence of bending and ripples in graphene modifies its electronic structure.\cite{cas09} Out-of-plane convex or concave deformations of the sheet have in principle equal probability and transitions between these deformations are associated with the bending energy of the sheet. A simple way to model this situation to consider that out-of-plane deformations are described by the values of an Ising spin associated to each carbon atom. We shall assume that the spins interact with each other only through their coupling with the membrane and have an energy 
\begin{eqnarray}
F_s= - f \int \sigma(x,y)\,w(x,y)\, dx dy, \label{eq3}
 \end{eqnarray}
where $f$ has units of force per unit area. At any time $t$, a spin located at $(x,y)$ may change its sign according to Glauber's dynamics: the atom-spin system may experience a transition from $(\bm{u},\bm{v},\bm{w},\bm{\sigma})$ to $(\bm{u},\bm{v},\bm{w},R_{(x,y)}\bm{\sigma})$ at a rate given by  \cite{Gl63}
\begin{eqnarray}\label{eq4}
&& W_{(x,y)}(\bm{\sigma}|\bm{u},\bm{v},\bm{w})=\frac{\alpha}{2} \left[1-\beta(x,y)\,\sigma(x,y)\right] , \\ 
&&\beta(x,y)=\tanh\left( \frac{f a^2w(x,y)}{\theta}\right)\!.  \label{eq5}
\end{eqnarray}
$R_{(x,y)}\bm{\sigma}$ is the configuration obtained from $\bm{\sigma}=\{\sigma(x,y)\}$ (for all points $(x,y)$ on the hexagonal lattice) by flipping the spin at the lattice point $(x,y)$. The parameter $\alpha$ gives the characteristic attempt rate for the transitions in the Ising system. Since the bending energy of the graphene sheet is $\kappa$, the attempt rate should be proportional to an Arrhenius factor, $\alpha=\alpha_0 e^{-\kappa/\theta}$,
where $\alpha_0$ is a constant. At room temperature, this is a large factor, $10^{-15}$, which bridges the gap between the phonon time scales and the observed damping time for lattice defects (a few seconds \cite{mey08}). 

The total free energy $F=F_g+F_s$ provides the equations of motion
 \begin{eqnarray}
 \rho_2 \partial_t^2 u&=& \lambda\,\partial_x\left( \partial_x u + \partial_y v+\frac{|\nabla w|^2}{2}\right) + \mu\,\partial_x [2 \partial_x u + (\partial_xw)^2]\nonumber\\
 &+& \mu\,\partial_y\left( \partial_y u + \partial_x v+ \partial_xw\partial_y w \right), \label{eq6}\\
 \rho_2 \partial_t^2 v &=& \lambda\,\partial_y\left( \partial_x u + \partial_y v+\frac{|\nabla w|^2}{2}\right) + \mu\,\partial_y[2 \partial_y v + (\partial_yw)^2]\nonumber\\
 &+& \mu\,\partial_x\left( \partial_y u + \partial_x v+ \partial_xw\partial_y w \right),  \label{eq7}
  \end{eqnarray}
\begin{eqnarray}
&&  \rho_2 \partial_t^2w =\lambda\,\partial_x\left[\left( \partial_x u + \partial_y v+\frac{|\nabla w|^2}{2}\right)\partial_xw\right]\nonumber\\
&&+ \lambda\,\partial_y\left[\left( \partial_x u + \partial_y v+\frac{|\nabla w|^2}{2}\right)\partial_yw\right]\nonumber\\
&&+ \mu\,\partial_x\left\{2 \partial_x u\partial_xw +(\partial_y u + \partial_x v)\partial_yw + |\nabla w|^2\partial_xw \right\} \nonumber\\
&&+ \mu\,\partial_y\left\{( \partial_y u + \partial_x v)\partial_xw+ 2\partial_y v\partial_y w\right.\nonumber\\
&&\quad \left. + |\nabla w|^2\partial_yw\right\}-\kappa\, (\nabla^2)^2w+f\sigma
+ \nabla\cdot(P\,\nabla w), \label{eq8}
 \end{eqnarray}
where $\rho_2$ is the 2D mass density (mass per unit area). The forcing term $f\sigma$ in the right hand side (RHS) of (\ref{eq8}) tries to bend the membrane upwards or downwards depending on the value of the spin at the point $(x,y)$. $\sigma(x,y)$ is a stochastic variable that flips with the transition probability (\ref{eq4}). The term $\nabla\cdot( P\,\nabla w)$ included in the RHS of (\ref{eq8}) represents the effect of a stress $P$ on the membrane. A homogeneous strain $u_{xx}^*=\gamma$, $u_{yy}^*=-\lambda\gamma/(\lambda+2\mu)$, $u_{xy}^*=0$, in a rectangular membrane of length $L$ and width $W$ elongated in the $x$ direction and pinned at $x=0$ and $x=L$ produces a stress $P_0=E_2\gamma$, where $E_2=4\mu(\lambda+\mu)/(\lambda+2\mu)$ is the 2D Young's modulus. In addition, the membrane may be deflected vertically which yields an additional stress $\Delta P=E_2\Delta S/S$, where $\Delta S$ is the change in the membrane area $S= LW$ due to bending:
\begin{eqnarray}
\frac{\Delta S}{S}=\int_0^L\int_0^W\frac{\sqrt{1+|\nabla w|^2}-1}{LW}\, dxdy\approx \frac{1}{2LW}\int_0^L\int_0^W|\nabla w|^2 dxdy.\nonumber
\end{eqnarray}
Then
 \begin{eqnarray}
P=\frac{4\mu(\lambda+\mu)}{\lambda+2\mu}\left(\gamma+ \frac{1}{2LW}\int_0^L\int_0^W|\nabla w|^2 dxdy\right)\!.\label{eq9}
\end{eqnarray}

\section{Equilibrium configurations and buckling phase transition}
\label{sec:3}
\subsection{Effective free energy provided by spins}
In equilibrium the part of the partition function due to spins is 
\begin{eqnarray}
Z_{\rm Ising}=\sum_{\bm{\sigma}} \exp\left[\frac{f\int\sigma(x,y)w(x,y)\, dxdy}{k_BT}\right] \nonumber\\
= \sum_{\bm{\sigma}} \exp\left[\frac{fa^2\sum_{x,y}\sigma(x,y)w(x,y)}{\theta}\right]=
\sum_{\bm{\sigma}}\prod_{x,y} \exp\left[\frac{fa^2\sigma(x,y)w(x,y)}{\theta}\right] \nonumber\\
= \prod_{x,y} 2\cosh\left[\frac{fa^2w(x,y)}{\theta}\right] = \exp\sum_{x,y}\ln\left\{2\cosh\left[\frac{fa^2w(x,y)}{\theta}\right]\right\} \nonumber\\
=\exp\left\{\frac{1}{a^2}\int\ln\left[2\cosh\left(\frac{fa^2w(x,y)}{\theta}\right)\right]dxdy\right\} \!. \nonumber
\end{eqnarray}
To obtain this equation we have replaced $\int (\ldots)\, dx dy$ by $a^2\sum_{x,y}$ wherever convenient. Added to (\ref{eq1}), the previous equation gives an additional contribution to the free energy of the membrane so that its effective free energy is
\begin{eqnarray}
&& F_{\rm eff}\!=\!\!\int\!\!\left\{\!\frac{\kappa}{2}(\nabla^2w)^2\! + \frac{\lambda}{2} u_{ii}^2 + \mu u_{ik}^2\!- \frac{\theta}{a^2}\ln\!\!\left[2\cosh\!\left(\frac{fa^2w(x,y)}{\theta}\right)\!\right]\!\right\} dx\, dy.  \label{eq10}
\end{eqnarray}
The equilibrium configuration is obtained by finding the minimum of $F_{\rm eff}$ with respect to the displacement vector. The components $u$, $v$ satisfy (\ref{eq6}) and (\ref{eq7}) with zero inertia ($\rho_2=0$), whereas (\ref{eq8}) should be replaced by
\begin{eqnarray}
&&\lambda\,\partial_x\left[\left( \partial_x u + \partial_y v+\frac{|\nabla w|^2}{2}\right)\partial_xw\right]+ \lambda\,\partial_y\left[\left( \partial_x u + \partial_y v+\frac{|\nabla w|^2}{2}\right)\partial_yw\right]\nonumber\\
&&+ \mu\,\partial_x\left\{2 \partial_x u\partial_xw +(\partial_y u + \partial_x v)\partial_yw%\right.\nonumber\\&&\quad \left. 
+ |\nabla w|^2\partial_xw \right\} \nonumber\\
&&+ \mu\,\partial_y\left\{( \partial_y u + \partial_x v)\partial_xw+ 2\partial_y v\partial_y w+ |\nabla w|^2\partial_yw\right\}\nonumber\\
&& -\kappa\, (\nabla^2)^2w+f\tanh\left(\frac{fa^2w}{\theta}\right) + P\nabla^2 w=0, \label{eq11}
 \end{eqnarray}
 The flat sheet $u=v=w=0$ is obviously a solution. Linearization of (\ref{eq11}) about it leads to the equation 
 \begin{eqnarray}
-\kappa\, (\nabla^2)^2\delta w+\frac{f^2a^2}{\theta}\delta w + P_0\nabla^2 \delta w=0, \label{eq12}
 \end{eqnarray}
where $P_0=E_2\gamma$. The flat sheet solution will lose stability at temperatures at which Equation (\ref{eq12}) has a nonzero solution. Let us assume that the rectangular sheet is simply supported at all its boundary points. Then we have $\delta w= \partial_x^2\delta w=0$ at $x=0$ and $x=L$ and $\delta w=\partial_y^2\delta w=0$ at $y=0,\, W$, so that $\delta w^{(n,m)}=\sin (n\pi x/L)\sin (m\pi y/W)$, and therefore $[P_0 \nabla^2-\kappa\, (\nabla^2)^2] \delta w^{(n,m)}=-P_0\pi^2[(n/L)^2+(m/W)^2]-\kappa\pi^4[(n/L)^2+(m/W)^2]^2$. Substituting in (\ref{eq12}), we find the critical temperatures 
\begin{eqnarray}
\theta_{n,m}=\frac{f^2 a^2}{P_0\pi^2\!\left(\frac{n^2}{L^2}+\frac{m^2}{W^2}\right)\!+\kappa\pi^4\!\left(\frac{n^2}{L^2}+\frac{m^2}{W^2}\right)^2}. \label{eq13}
\end{eqnarray}
The maximum temperature $\theta_c=\theta_{1,1}$ gives the critical temperature above which the flat graphene sheet is stable. Below this temperature stable ripples form. The same result will be obtained below from the macroscopic equations for the average displacement vector.

\subsection{Macroscopic equations and critical temperature} 
In time-resolved experiments, each framework in a movie consists of data taken during one second \cite{mey08}. Correspondingly, the stochastic equations of motion (\ref{eq6})-(\ref{eq8}) with the Glauber dynamics (\ref{eq4})-(\ref{eq5}) have to be integrated over long time intervals and averaged over shorter intervals during one second. These time averages smooth out the displacements of the carbon atoms which become very close to their mean values. The equations for the average displacements obtained from (\ref{eq6})-(\ref{eq8}) have to be supplemented by the equation for $\langle\sigma(x,y)\rangle=\widetilde{q}(x,y)$ \cite{Gl63},
\begin{eqnarray}
\partial_t\widetilde{q} = \frac{\alpha}{2}\left[\left\langle\tanh\left( \frac{f a^2w(x,y)}{\theta}\right)\right\rangle-\widetilde{q}\right]. \label{eq14}
\end{eqnarray}
To analyze the average equations, we split $u=\widetilde{u}+\Delta u$, etc., in their averages $\widetilde{u}= \langle u\rangle$ and fluctuations $\Delta u$. For a large enough sheet, we ignore the fluctuations and then the average values satisfy  
\begin{eqnarray}
&& \rho_2 \partial_t^2 \widetilde{u}= \lambda\,\partial_x\left( \partial_x\widetilde{u} + \partial_y\widetilde{v}+\frac{|\nabla\widetilde{w}|^2}{2}\right) + \mu\,\partial_x [2 \partial_x\widetilde{u}+ (\partial_x\widetilde{w})^2]\nonumber\\
 &&\quad + \mu\,\partial_y\left( \partial_y\widetilde{u} + \partial_x \widetilde{v}+ \partial_x\widetilde{w}\partial_y \widetilde{w} \right), \label{eq15}\\
&& \rho_2 \partial_t^2\widetilde{v} = \lambda\,\partial_y\left( \partial_x\widetilde{u} + \partial_y\widetilde{v}+\frac{|\nabla\widetilde{u}|^2]}{2}\right) +\mu\,\partial_y[2 \partial_y \widetilde{v}+ (\partial_y\widetilde{w})^2] \nonumber\\
 &&\quad+ \mu\,\partial_x\left( \partial_y\widetilde{u} + \partial_x\widetilde{v}+ \partial_x\widetilde{w}\partial_y\widetilde{w} \right),  \label{eq16}\\
&&\rho_2 \partial_t^2\widetilde{w}= \lambda\,\partial_x\left[\left( \partial_x \widetilde{u} + \partial_y \widetilde{v}+\frac{|\nabla\widetilde{w}|^2}{2}\right)\partial_x\widetilde{w}\right]\nonumber\\
&& +\lambda\,\partial_y\left[\left( \partial_x\widetilde{u} + \partial_y\widetilde{v}+\frac{(\partial_x\widetilde{w})^2 + (\partial_y\widetilde{w})^2}{2}\right)\partial_y\widetilde{w}\right]\nonumber\\
&&+ \mu\,\partial_x\left\{2 \partial_x u\partial_x\widetilde{w} +(\partial_y\widetilde{u} + \partial_x\widetilde{v})\partial_y\widetilde{w}+ |\nabla\widetilde{w}|^2 \partial_x\widetilde{w} \right\} \nonumber\\
&&+ \mu\,\partial_y\left\{( \partial_y\widetilde{u} + \partial_x\widetilde{v})\partial_x\widetilde{w}+ 2\partial_y\widetilde{v}\partial_y\widetilde{w}+  |\nabla\widetilde{w}|^2\partial_y\widetilde{w}\right\}\nonumber\\
&&-\kappa\, (\nabla^2)^2\widetilde{w}+f\widetilde{q}+\frac{4\mu(\lambda+\mu)}{\lambda+2\mu}\!\left(\gamma+ \frac{1}{2LW}\int_0^L\int_0^W|\nabla\widetilde{w}|^2 dxdy\right)\!\nabla^2\widetilde{w}, \label{eq17}\\
&&\partial_t\widetilde{q} = \frac{\alpha}{2}\left[\tanh\left( \frac{f a^2\widetilde{w}(x,y)}{\theta}\right)-\widetilde{q}\right]. \label{eq18}
\end{eqnarray}
These equations have the trivial solution $\widetilde{u}=\widetilde{v}=\widetilde{w}=\widetilde{q}=0$ (flat sheet). Separation of variables in the linear stability problem ($\widetilde{u}=e^{\Lambda t}\psi_u$, \ldots) leads to the eigenvalue equations ($P_0=E_2\gamma$)
\begin{eqnarray}
&& \rho_2 \Lambda^2 \psi_{u}= (\lambda+2\mu)\partial_x^2\psi_{u} + (\lambda+\mu)\partial_{xy}^2\psi_{v} + \mu\partial_y^2 \psi_{u},\quad\, \label{eq19}\\
&& \rho_2 \Lambda^2\psi_{v} = (\lambda+\mu)\partial^2_{xy}\psi_{u} + (\lambda+2\mu)\partial_y^2\psi_{v}+\mu\partial_x^2\psi_{v},\quad\,  \label{eq20}\\
&&\rho_2 \Lambda^2\psi_{w}= f\psi_{q}+P_0\nabla^2\psi_{w}-\kappa\, (\nabla^2)^2\psi_{w}, \label{eq21}\\
&&\Lambda\psi_{q} = \frac{\alpha}{2}\left( \frac{f a^2\psi_{w}}{\theta}-\psi_{q}\right). \label{eq22}
\end{eqnarray}
Eliminating $\psi_q$ from the last two equations, we obtain 
 \begin{eqnarray}
\rho_2 \Lambda^2\psi_{w}&=& P_0\nabla^2\psi_{w}-\kappa\, (\nabla^2)^2\psi_{w}+\frac{f^2 a^2\,\psi_{w}}{\left(1+\frac{2\Lambda}{\alpha}\right)\theta}. \label{eq23}
\end{eqnarray}

As in the previous section, we shall assume that the rectangular sheet is simply supported at all its boundary points. We again get eigenfunctions $\psi_w^{(n,m)}=\sin (n\pi x/L)\sin (m\pi y/W)$, with $[P_0\nabla^2-\kappa\, (\nabla^2)^2] \psi_{w}^{(n,m)}=-P_0 \pi^2[(n/L)^2+(m/W)^2]-\kappa\pi^4[(n/L)^2+(m/W)^2]^2$. Substituting in (\ref{eq23}), we find the eigenvalue equation
\begin{eqnarray}
\rho_2 \Lambda^2+P_0\pi^2\!\left(\frac{n^2}{L^2}+\frac{m^2}{W^2}\right)\!+\kappa\pi^4\!\left(\frac{n^2}{L^2}+\frac{m^2}{W^2}\right)^2 -\frac{f^2 a^2}{\left(1+\frac{2\Lambda}{\alpha}\right)\theta}=0. \label{eq24}
\end{eqnarray}
We have $\Lambda= 0$ at the temperatures $\theta_{n,m}$ given by (\ref{eq13}). The largest such temperature is $\theta_c=\theta_{1,1}$:
\begin{eqnarray}
\theta_{c}=\frac{1}{\pi^2}\frac{f^2 a^2L^4}{P_0L^2\!\left(1+\frac{L^2}{W^2}\right)\!+\kappa\pi^2\!\left(1+\frac{L^2}{W^2}\right)^2}. \label{eq25}
\end{eqnarray}
Above $\theta_c$, the flat graphene sheet is stable and spins are unpolarized, whereas long non-homogeneous ripple patterns with wavelengths related to $L$ and $W$ are formed below $\theta_c$ and the spins become polarized. Note that increasing the strain $\gamma$ decreases $\theta_c$ which favors ripple destruction. The spins phase transition at $\theta_c$ is related to a buckling transition in the graphene sheet: The stable state for $\theta$ just below $\theta_c$ corresponds to a curved sheet all whose atoms are above or below the horizontal plane. As $\theta$ further decreases, other critical temperatures $\theta_{n,m}$ are reached for which patterns corresponding to eigenfunctions $\psi_w^{(n,m)}$ with $n-1$ (resp.\ $m-1$) zeros in the horizontal (resp.\ vertical) direction may appear. These patterns are long ripples in the graphene sheet. For a simpler model of a harmonic oscillator coupled to Ising spins undergoing Glauber dynamics, the connection between the spins phase transition at $\theta_c$ and the instability of the rest state of the oscillator and its dynamics are discussed in \cite{prados,BCP10}. The case of rippling in a chain of oscillators each connected to an Ising spin undergoing Glauber dynamics is considered in \cite{bcpr11}. Similar systems with other short-range oscillator-spin couplings were studied in relation with the collective Jahn-Teller effect \cite{pyt73,rik77}. Those early studies were focused on whether spin degeneracy is broken below a certain critical temperature and on the spin dynamics, but they did not consider the possible patterns of the oscillator systems as \cite{bcpr11} and the present work do.

\section{Numerical simulations of periodized discrete elasticity.} 
\label{sec:4}
\begin{figure}
\begin{center}
\includegraphics[width=8cm]{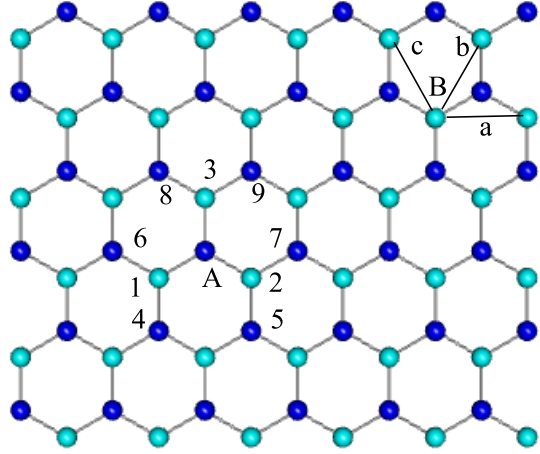}
\caption{(Color online) Neighbors of a given atom $A$ in sublattice 1 (dark blue). }
\label{fig1}
\end{center}
\end{figure}
We discretize the equations of motion on a hexagonal lattice using the same notation as in \cite{car08,CBJV08,bon11}. Assigning the coordinates $(x,y)$ to the atom $A$ in sublattice 1 as in Figure \ref{fig1}, the three nearest neighbors of $A$ belong to sublattice 2 and their cartesian coordinates are $n_1$, $n_2$ and $n_3$ below. Its six next-nearest neighbors belong to sublattice 1 and their cartesian coordinates are $n_i$, $i=4,\ldots, 9$:
\begin{eqnarray}
&& n_{1}=\left(x-\frac{a}{ 2},y-\frac{a}{ 2\sqrt{3}}\right)\!\!, \, n_{2}=\left(x+\frac{a}{ 2}, y-\frac{a}{ 2\sqrt{3}}\right)\!\!, \nonumber\\ 
&& n_{3}=\left(x,y+\frac{a}{\sqrt{3}}\right)\!,\quad n_{4}=\left(x-\frac{a}{ 2},y-\frac{a\sqrt{3}}{ 2}\right)\!,\nonumber\\
&& n_{5}=\left(x+\frac{a}{ 2},y-\frac{a\sqrt{3}}{ 2}\right)\!, \quad n_{6}=(x-a,y), \nonumber\\ 
&& n_{7}=(x+a,y),\quad n_{8}=\left(x-\frac{a}{ 2},y+\frac{a\sqrt{3}}{ 2}\right)\!,\nonumber\\
&& n_{9}=\left(x+\frac{a}{ 2},y+\frac{a\sqrt{3}}{ 2}\right)\!. \label{eq26}
\end{eqnarray}
In Fig. \ref{fig1}, atoms $n_6$ and $n_7$ are separated from $A$ by the primitive vector $\pm {\bf a}$ and atoms $n_4$ and $n_9$ are separated from $A$ by the primitive vector $\pm {\bf b}$. Instead of choosing the primitive vector $\pm {\bf b}$, we could have selected the primitive direction $\pm {\bf c}$ along which atoms $n_8$, $A$ and $n_5$ lie. Let us define the following operators acting on functions of the coordinates $(x,y)$ of node $A$:
\begin{eqnarray}
Tu &=& [u(n_{1})-u(A)] + [u(n_{2})-u(A)] \nonumber\\
&+& [u(n_{3})-u(A)]\sim (\partial_x^2 u + \partial_y^2 u)\,\frac{a^2}{4},\label{eq27}\\
Hu &=& [u(n_{6})-u(A)]+[u(n_{7})-u(A)]\sim a^2\partial_x^2 u,\,\,  \label{eq28} \\
D_{1}u &=& [u(n_{4})-u(A)] + [u(n_{9})-u(A)]  \label{eq29}\\
&\sim& \left(\frac{1}{ 4}\, \partial_x^2 u +\frac{\sqrt{3}}{ 2}\,\partial_x\partial_yu + \frac{3}{ 4}\, \partial_y^2 u\right)a^2,\nonumber\\
D_{2}u &=& [u(n_{5})-u(A)] + [u(n_{8})-u(A)]  \label{eq30}\\
&\sim&\left(\frac{1}{ 4}\, \partial_x^2 u -\frac{\sqrt{3}}{ 2}\,\partial_x\partial_yu + \frac{3}{ 4}\, \partial_y^2 u\right)a^2,  \nonumber\\
\Delta_hu &=& u(n_{7})-u(A)\sim (\partial_xu)\, a, \label{eq31}\\
\Delta_v u &=& u(n_{3})-u(A)\sim (\partial_yu)\, \frac{a}{\sqrt{3}}, \label{eq32}\\
Bw&=& [Tw(n_1) - Tw(A)] + [Tw(n_2) - Tw(A)] \nonumber\\
&+& [Tw(n_3) - Tw(A)]\sim \frac{a^4}{16} (\partial_x^2+\partial_y^2)^2w.\label{eq33}
\end{eqnarray}
with similar definitions for points $A$ in sublattice 2. Taylor expansions of these finite difference combinations ($T$, $H$, $D_1$ and $D_2$ were defined in \cite{car08}) about $(x,y)$ yield the partial derivative expressions written above as $a\to 0$. We now replace $Hu/a^2$, $(4T-H)u/a^2$ and $(D_{1}-D_{2})u/(\sqrt{3}a^2)$, $\Delta_hw/a$, $\sqrt{3}\Delta_vw/a$ and $16 Bw/a^4$ instead of $\partial_x^2 u$, $\partial_y^2 u$, $\partial_x\partial_yu$, $\partial_x w$, $\partial_yw$ and $(\nabla^2)^2w$, respectively, in (\ref{eq6}), (\ref{eq7}) and (\ref{eq8}) with similar substitutions for the derivatives of $v$ and $w$. We obtain the following equations at each point of the lattice:
\begin{eqnarray}
\rho_2 a^2 \partial_t^2 u &=& 4\mu\, Tu + (\lambda+\mu)\, Hu +\frac{\lambda+\mu }{\sqrt{3}}\, (D_{1} - D_{2})v  \nonumber\\
&+&\frac{\lambda+\mu}{a}\, [\Delta_hw\, Hw+\Delta_vw\, (D_1-D_2)w]+ \frac{4\mu}{a}\, \Delta_hw\, Tw, \label{eq34}\\
\rho_2 a^2 \partial_t^2 v &=& 4 (\lambda+2\mu )\, Tv +\frac{\lambda+\mu}{\sqrt{3}}\, (D_{1} - D_{2})u \nonumber\\
&-& (\lambda+\mu )H v+ \frac{4\sqrt{3}}{a} (\lambda+2\mu)\Delta_vw\, Tw\nonumber\\
&+&\frac{\lambda+\mu}{a\sqrt{3}}\, [\Delta_hw\, (D_1-D_2)w-3\Delta_vw\, Hw], \label{eq35}
\end{eqnarray}
\begin{eqnarray}
\rho_2 a^2\partial_t^2w &=& \frac{\lambda+2\mu}{a}\left\{\left[Hu+\frac{2\Delta_vw}{a}(D_1-D_2)w+\frac{\Delta_hw}{a}Hw\right]\Delta_hw 
\right.\nonumber\\
&+&\left. \left[ \sqrt{3}(4T-H)v+\frac{3\Delta_vw}{a}(4T-H)w\right]\Delta_vw\right\}\nonumber\\
%&+&\left.\left.\nonumber\\
&+& \frac{\lambda+\mu}{a}(D_1-D_2)u\Delta_vw+\frac{\lambda\Delta_hw}{\sqrt{3}a}(D_1-D_2)v\nonumber\\
&+&\frac{\mu\Delta_vw}{a}[\sqrt{3}(4T-H)u+\sqrt{3}Hv+(D_1-D_2)v] \nonumber\\
&+&\frac{2\Delta_hu}{a}(2\lambda T+\mu H)w+\frac{2\sqrt{3}\Delta_vv}{a}[2\lambda Tw+\mu (4T-H)w] \nonumber\\
&+& 2\mu\!\left(\!\Delta_v u\!+\!\frac{\Delta_h v}{\sqrt{3}}\!\right)\!\frac{(D_1\!-\!D_2)w}{a}%\nonumber\\&+&
+\frac{(\Delta_hw)^2\!+\!(\Delta_vw)^2\!}{a^2}(2\lambda T+\mu H)w\nonumber\\
&+&\frac{4\mu}{a}Tw(4T-H)w-\frac{16\kappa}{a^2} Bw+ 4P\, Tw+fa^2\sigma.\label{eq36}
\end{eqnarray}
Here
\begin{eqnarray}
P=\frac{4\mu(\lambda+\mu)}{\lambda+2\mu}\!\left(\gamma+\frac{1}{2Na^2}\sum_{x,y}[(\Delta_hw)^2+9(\Delta_vw)^2] \right)\!, \label{eq37}
\end{eqnarray}
and $N=\sum_{x,y} 1$ is the total number of atoms in the graphene sheet. The spin variable $\sigma$ stochastically flips with probability (\ref{eq4}). Possible defects inserted in the graphene sheet are the cores of dislocations. To account for them, we have to write these equations of motion in primitive coordinates and periodize all difference operators appearing in them along  primitive directions. \textit{Note that we can use the equations of motion (\ref{eq34})-(\ref{eq36}) without periodization if there are no defects in the graphene lattice}. These equations of motion become those in \cite{car08} for $w=0$. 

\begin{table}[ht]
\begin{center}\begin{tabular}{ccccc}
 \hline
$\frac{fa}{\lambda+2\mu}$  &$\frac{\theta}{fa^3}$ & $\frac{\mu}{\lambda+2\mu}$ & $\frac{16\kappa}{(\lambda+2\mu)a^2}$& $\delta=\frac{\alpha a\sqrt{\rho}_2}{\sqrt{\lambda+2\mu}}$ \\
$4.141\times 10^{-4}$ & $0.4593$&0.4428& 0.1271 & $4.084\times 10^{-16}$\\
 \hline
\end{tabular}
\end{center}
\caption{Dimensionless parameters corresponding to $f=3.78$ meV/\AA$^3$.}
\label{t1}
\end{table}

\begin{figure}
\begin{center}
\includegraphics[width=14cm]{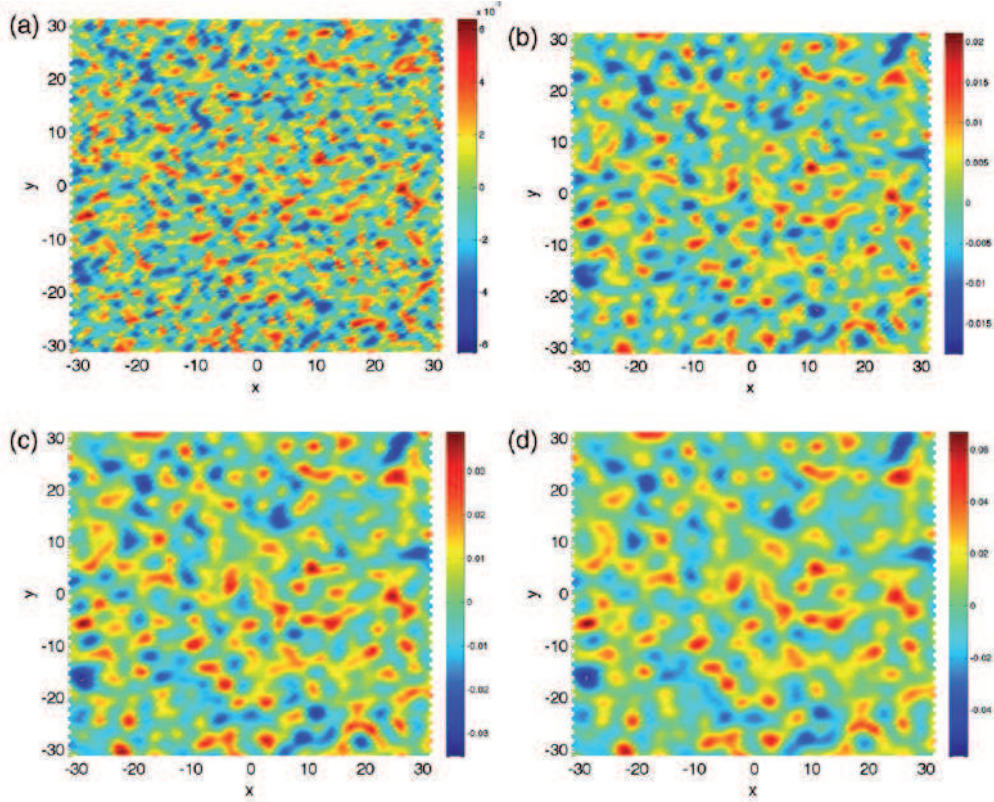}
\caption{(Color online) Ripples formed in a suspended $80\times 80$-hexagon graphene sheet after (a) 5, (b) 10, (c) 15, and (d) 20 nondimensional time units. Parameter values as that in Table \ref{t1} but with $\delta=\alpha a\sqrt{\rho}_2/\sqrt{\lambda+2\mu} =10^{-11}$. }
\label{fig2}
\end{center}
\end{figure}

\begin{figure}
\begin{center}
\includegraphics[width=14cm]{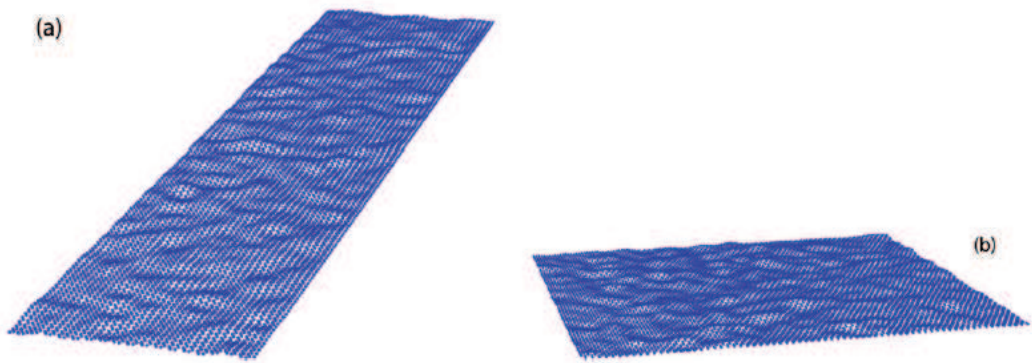}
\caption{(Color online) (a) Ripples formed in a suspended $160\times 40$-hexagon graphene sheet. (b) Same for a $80\times 80$-hexagon sheet. Parameter values are from \cite{zak09} at 300K with $f=0.378$ meV/\AA$^3$ (ten times smaller than that in Table \ref{t1}). $\delta=\alpha a\sqrt{\rho}_2/\sqrt{\lambda+2\mu} =10^{-4}$. }
\label{fig3}
\end{center}
\end{figure}

In our numerical calculations, we have used the values of the 2D Lam\'e moduli at 300K given in \cite{zak09} (that agree with experimental measurements \cite{bao09}), $\lambda+2\mu=22.47$ eV/\AA$^2$, $\mu=9.95 $ eV/\AA$^2$, $\kappa=1.08$ eV (see Figure 7 of \cite{zak10}) and $1/\alpha=27$ s (see \cite{car08}). To  calculate $f$, we balance the energy of a ripple of height $w_0$, $E_2 w_0\lambda_r$ ($\lambda_r$ is the ripple wavelength and $E_2=4\mu(\lambda+\mu)/(\lambda+2\mu)$ the 2D Young modulus), with the coupling energy due to spins, $fw_0LW$. We find $f= E_2 \lambda_r/(LW)$. For a unstrained small sample with $L=W=80a$ and $\lambda_r= 6.6$ \AA, $f=0.00378$ eV/\AA$^3$, the critical temperature is 11900 times the ambient temperature, and it is 9160 times the ambient temperature if there is a 1.5\% strain. The critical temperature is much larger for typical samples used in the experiments. Thus all graphene samples are in a subcritical buckled state.

Dimensionless parameters for our model have the numerical values listed in Table \ref{t1}. $u$, $v$, $w$ are zero for all atoms at the boundaries and for the atoms outside the sheet whose displacements enter (\ref{eq6})-(\ref{eq8}). In our tests, the average value of $|\nabla w|$ is between 0.1 and 0.001 and it increases with $f$. The time scale depends on $\alpha$. To speed up simulations, we have used a larger $\alpha$ (and therefore a larger $\delta= \alpha a\sqrt{\rho}_2/\sqrt{\lambda+2\mu}$) than in Table \ref{t1}. The results are similar if we use smaller $\alpha$, except that it takes more time to arrive at the rippled states shown in the figure. Figure \ref{fig2} shows the ripples formed in a square suspended graphene sheet having $80\times 80$ hexagons (12800 atoms) for different times. Initially the sheet is planar and the spins are in a random configuration. Figures \ref{fig2}(a) and (b) show that there appear small domains in which all atoms are displaced vertically upward or downward from the horizontal. These are the ripples. After a short transient, they become larger and taller and change very little as time evolves (cf. Figures \ref{fig2}(c) and (d)), only atoms in the periphery of a ripple domain flip. Figures \ref{fig3}(a) and (b) show ripples in rectangular ($160\times 40$ hexagons i.e., 12800 atoms) and square  ($80\times 80$ hexagons, also 12800 atoms) suspended graphene sheets. The dominant ripple wavelength is about 20 lattice spacings (5 nm) in short samples of $80\times 80$ spacings and it goes up to about 45 spacings (11 nm) for longer $160\times 160$ samples. There are ripples of shorter wavelength superimposed to the longer ones. Over long periods of time, significantly larger than the mean time between spin flips, the ripples change. This scenario agrees with the observations of Bangert et al \cite{ban09}.

Spontaneous ripple generation is caused by coupling with Ising spins. If $f=0$ (no coupling), the number and appearance of ripples depends on the initial condition. If the initial condition compatible with the boundary conditions contains a given number of ripples, they persist for subsequent times under the equations of periodized discrete elasticity. Adding tension $P$ flattens the ripples somewhat without suppressing them. When coupling with Ising spins is restored, ripples such as those in Figures \ref{fig2} and \ref{fig3} appear after a transient stage even from an initially random state. The average ripple height increases with increasing $f$ and decreasing $\theta$. For the boundary conditions we use, tension has little effect on the ripples.

\begin{figure}
\begin{center}
\includegraphics[width=14cm]{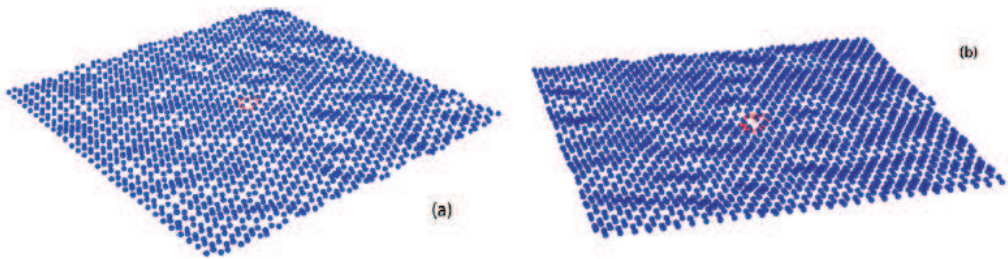}
\caption{(Color online) (a) Ripples formed in a suspended $40\times 40$-hexagon square graphene sheet containing a vacancy. (b) As in (a) but containing a divacancy. Parameters are as in Figure \ref{fig3} except that $\delta=10^{-6}$. }
\label{fig4}
\end{center}
\end{figure}

Figure \ref{fig4} shows rippled states for suspended sheets that contain stable defects in them: a vacancy in Fig. \ref{fig4}(a) and a divacancy or pentagon-octagon-pentagon defect in Fig.Ê\ref{fig4}(b). Both these defects are the cores of dislocation dipoles of equal and opposite Burgers vectors and we have calculated them by extending our previous procedure from planar graphene \cite{car08} to the present model that allows off plane displacements. Vacancies and divacancies seem to surf on the ripples  affecting little the local curvature of the graphene sheet in seeming agreement with experiments \cite{ban09}. In contrast to this, a single pentagon-heptagon effect (dislocation with nonzero Burgers vector) strongly affects the sheet local curvature: The sheet is pushed upwards near the pentagon and downwards near the heptagon. These effects are due to the coupling with the spins: there is no systematic bending of the graphene sheet near defects if $f=0$ and ripples are introduced through an initial condition. In this case, the initial condition determines the local bending of the sheet near defects.

We have also explored other couplings between spins and carbon atoms. For instance we could have coupled the spins to the sheet curvature thereby choosing
\begin{eqnarray}
F_s= - fa^2 \int \sigma(x,y)\,\nabla^2 w(x,y)\, dx dy, \label{eq38}
\end{eqnarray}
instead of (\ref{eq3}). This is similar to the electron-phonon coupling in \cite{sgg11}. In this case the sheet also buckles upward or downward for temperatures below a certain critical value. However, simulations of this model show buckling but no ripple formation.

\section{Conclusions} 
\label{sec:5}
We have proposed a mechanism to spontaneously produce ripples in a suspended graphene sheet. Besides the membrane free energy, carbon atoms are coupled to Ising spins at each lattice point that flip randomly according to Glauber dynamics. The spins generate forces that pull atoms out from the planar configuration of the graphene sheet. After a short transient, there appear domains in which all atoms are vertically displaced above or below the horizontal plane. These domains acquire a certain size and form ripples with no preferred direction. Changes to the domains are slow and are produced only through changes in the atoms at their boundaries. When the spin flipping rate is much lower than the frequencies of the membrane, the ripples are long-lived metastable states for any temperature. Numerical simulations show that, after a transient stage, stable rippling appears even from random initial states. If there is no coupling to spins, ripples exist only if they are generated by initial conditions. Pentagon-heptagon defects corresponding to dislocations with nonzero Burgers vector strongly affect the local curvature of the sheet whereas defects such as vacancies or divacancies (corresponding to dislocation dipoles with zero overall Burgers vector) surf on the ripples. 

\ack
This work has been supported by the Spanish Ministerio de Econom\'\i a y Competitividad grants FIS2011-28838-C02-01,  FIS2011-28838-C02-02 and FIS2010-22438-E (Spanish National Network Physics of Out-of-Equilibrium Systems). The authors thank M.P. Brenner for hospitality during a stay at Harvard University financed by Fundaci\'on Caja Madrid mobility grants.


\section*{References}
\begin{thebibliography}{28}
\bibitem{nov04}
Novoselov K S, Geim A K, Morozov S V, Jiang D, Zhang Y, Dubonos S V, Grigorieva I V,  and Firsov A A, 2004 \textit{Science} {\bf 306} 666
\bibitem{nov05}
Novoselov K S, Jiang D, Schedin F, Booth T J, Khotkevich V V, Morozov S V and Geim A K, 2005 \textit{Proc. Natl. Acad. Sci. USA} {\bf 102} 10451 
\bibitem{gei07} 
Geim A K and Novoselov K S, 2007 \textit{Nature Materials} {\bf 6} 183 
\bibitem{gei09}
Geim A K, 2009 \textit{Science} {\bf 324} 1530 
\bibitem{cas09}
Castro Neto A H, Guinea F, Peres N M R, Novoselov K S and Geim A K, 2009 \textit{Rev. Mod. Phys.} {\bf 81} 109 
\bibitem{voz10}
Vozmediano M A H, Katsnelson M I and Guinea F, 2010 \textit{Phys. Rep.} {\bf 496} 109 
\bibitem{mey07} 
Meyer J C, Geim A K, Katsnelson M I, Novoselov K S, Booth T J and Roth S, 2007 \textit{Nature} {\bf 446} 60 
\bibitem{ban09} 
Bangert U, Gass M H, Bleloch A L, Nair R R and Geim A K, 2009 \textit{Physica status solidi (a)} {\bf 206}, 1117
\bibitem{gei08}
Katsnelson M I and Geim A K, 2008 \textit{Phil. Trans. R. Soc. A} {\bf 366} 195 
\bibitem{sgg11}
San-Jose P, Gonz\'alez J and Guinea F, 2011 \PR Lett. {\bf 106} 045502 
\bibitem{kim08}
Kim E A and Castro Neto A H, 2008 \textit{Europhys. Lett.} {\bf 84} 57007
 \bibitem{gaz09}
Gazit D, 2009 \PR B {\bf 80} 161406(R) 
\bibitem{fas07} 
Fasolino A, J.H. Los J H and Katsnelson M I, 2007 \textit{Nature Materials} {\bf 6} 858 
\bibitem{abe07}
Abedpour N, Neek-Amal M, Asgari R, Shahbazi F, Nafari N and Tabar M R, 2007 \PR B {\bf 76} 195407 
\bibitem{tho09} 
Thompson-Flagg R C, Moura M J B and Marder M, 2009 \textit{Europhys. Lett.} {\bf 85} 46002 
\bibitem{gui09}
Guinea F, Katsnelson M I and Geim A K, 2009 \textit{Nature Phys.} {\bf 6}, 30 
\bibitem{bao09}
Bao W, Miao F, Chen Z, Zhang H, Jang W, Dames C and Lau C N, 2009 \textit{Nature Nanotech.} {\bf 4} 562 
\bibitem{mao11}
Mao Y, Wang W L, Wei D, Kaxiras E and Sodroski J G, 2011 \textit{ACS Nano} {\bf 5} 1395
\bibitem{cer03} Cerda E and Mahadevan L, 2003 \PR Lett. {\bf 90} 074302
\bibitem{oha12}
O'Hare A, Kursmartsev F V and Kugel K I, 2012 \textit{Nano Lett.} {\bf 12} 1045 
\bibitem{Gl63}
Glauber R J, 1963 \textit{J. Math. Phys.} {\bf 4} 294
\bibitem{CN93}
Carraro C and Nelson D R, 1993 \PR E {\bf 48} 3082 
\bibitem{CN96}
Chou T and Nelson D R, 1996 \PR E {\bf 53} 2560 
\bibitem{bcpr11}
Bonilla L L, Carpio A, Prados A and Rosales R R, 2012 \PR E {\bf 85} 031125 
\bibitem{CBJV08}
Carpio A, Bonilla L L, de Juan F and Vozmediano MAH, 2008 \textit{New J. Phys.} {\bf 10} 053021
\bibitem{car08}
Carpio A and Bonilla L L, 2008 \PR B {\bf 78} 085406 
\bibitem{bon11}
Bonilla L L and Carpio A, 2011 \textit{Continuum Mech. Thermodyn.} {\bf 23} 337 
\bibitem{BC11}
Bonilla L L and Carpio A, 2011 \textit{Graphene Simulation}, edited by J R Gong. ISBN: 978-953-307-556-3 (Rijeka, Croatia: Intech), pp 167-182
\bibitem{nel02} 
Nelson D R, 2002 \textit{Defects and Geometry in Condensed Matter Physics} (Cambridge: Cambridge University Press).
\bibitem{mey08}
Meyer J C, Kisielowski C, Erni R, Rossell M D, Crommie M F and Zettl A, 2008 \textit{Nano Lett.} {\bf 8} 3582 
\bibitem{prados} 
Prados A, Bonilla L L and Carpio A, 2010 \textit{J. Stat.  Mech.} P06016
\bibitem{BCP10}
 Bonilla L L, Prados A and Carpio A, 2010 \textit{J. Stat. Mech.} P09019 
\bibitem{pyt73}
Feder J and Pytte E, 1973 \PR B{\bf 8} 3978
\bibitem{rik77} Rikvold P A, 1977 \textit{Z. Phys.} B {\bf 26} 195
\bibitem{zak09} 
Zakharchenko K V, Katsnelson M I and Fasolino A, 2009 \PR Lett.  {\bf 102} 046808 
\bibitem{zak10}
Zakharchenko K V, Los J H, Katsnelson M I and Fasolino A, 2010 \PR B {\bf 81}, 235439 

\end{thebibliography}
\end{document}